\documentclass[aip, amsmath,amssymb,reprint, longbibliography, final]{revtex4-2}
\usepackage[colorlinks,bookmarks=false,citecolor=blue,linkcolor=red,urlcolor=blue]{hyperref}

\usepackage[usenames,dvipsnames]{color}
\usepackage{soul}

\usepackage{amsfonts}
\usepackage{amsmath}

\usepackage{graphicx}
\usepackage{dcolumn}
\usepackage{bm}

\usepackage[utf8]{inputenc}
\usepackage[T1]{fontenc}
\usepackage{mathptmx}

\usepackage[dvipsnames,table]{xcolor}

\usepackage{siunitx}
\mathchardef\mhyphen="2D
\usepackage{floatrow}
\usepackage{multirow}
\usepackage{upgreek}
\usepackage{makecell}
\usepackage{tabularx}
\usepackage{physics}

\newcommand{\suppl}[1]{Supplementary}
\newcommand{\eqnref}[1]{Eq.~\eqref{eq:#1}}

\newcommand{\tabref}[1]{Tab.~\ref{tab:#1}}

\newcommand{\subfigref}[2]{Fig.~\hyperref[fig:#1]{\ref*{fig:#1}#2}}
\newcommand{\Supplsubfigref}[2]{\suppl{}~Fig.~\hyperref[fig:#1]{\ref*{fig:#1}#2}}
\newcommand{\SupplsubfigrefL}[2]{\suppl{}~Figure~\hyperref[fig:#1]{\ref*{fig:#1}#2}}
\newcommand{\SupplsubfigrefS}[2]{Fig.~\hyperref[fig:#1]{\ref*{fig:#1}#2}}
\newcommand{\subfigrefL}[2]{Figure~\hyperref[fig:#1]{\ref*{fig:#1}#2}}
\newcommand{\subfigsref}[3]{Figs.~\hyperref[fig:#1]{\ref*{fig:#1}#2}-\hyperref[fig:#1]{\ref*{fig:#1}#3}}
\newcommand{\Supplsubfigsref}[3]{\suppl{}~Figs.~\hyperref[fig:#1]{\ref*{fig:#1}#2}-\hyperref[fig:#1]{\ref*{fig:#1}#3}}
\newcommand{\subfigsrefL}[3]{Figures~\hyperref[fig:#1]{\ref*{fig:#1}#2}-\hyperref[fig:#1]{\ref*{fig:#1}#3}}
\newcommand*{\figlab}[1]{\textbf{#1},}

\newcommand*{\fp}[1]{$F_{\mathrm{P}}$}
\newcommand*{\lowT}[1]{$T = \SI{4.2}{\kelvin}$}
\newcommand*{\methods}[1]{Methods}

\DeclareSIUnit\ML{ML}
\DeclareSIUnit\px{px}
\newcommand*{\SM}[1]{\suppl{}~Information}

\usepackage{array}
\usepackage{tabularx}
\usepackage{arraycols}

\usepackage[flushleft]{threeparttable}

\usepackage{booktabs}

\usepackage{etoolbox}
\robustify\bfseries
\robustify\itshape

\newcolumntype{E}{>{\centering\arraybackslash}X}

\usepackage{microtype}
\usepackage{rotating}
\usepackage{textcomp}

\newcommand*{\oldtext}[2][Old text: ]{\textit{#1}{\textcolor{Red}{#2}}}

\makeatletter
\let\@fnsymbol\@fnsymbol@latex
\@booleanfalse\altaffilletter@sw
\makeatother

\begin{document}
\preprint{AIP/123-QED}

\newcommand*{\papertitle}{Telecom-band site-controlled quantum dots with engineered low fine-structure splitting}

\author{Christian C. Ruiz Madera}
\affiliation{DTU Electro, Department of Electrical and Photonics Engineering, Technical University of Denmark, Ørsteds Plads 343, DK-2800 Kongens Lyngby, Denmark}
\affiliation{NanoPhoton - Center for Nanophotonics, Technical University of Denmark, Ørsteds Plads 345A, DK-2800 Kongens Lyngby, Denmark}

\author{Pawe\l{}~Holewa}
\email{pawel.holewa@pwr.edu.pl}
\affiliation{DTU Electro, Department of Electrical and Photonics Engineering, Technical University of Denmark, Ørsteds Plads 343, DK-2800 Kongens Lyngby, Denmark}
\affiliation{NanoPhoton - Center for Nanophotonics, Technical University of Denmark, Ørsteds Plads 345A, DK-2800 Kongens Lyngby, Denmark}
\affiliation{Department of Experimental Physics, Faculty of Fundamental Problems of Technology, Wrocław University of Science and Technology, Wyb. Wyspiańskiego 27, 50-370 Wrocław, Poland}

\author{Paweł Wyborski}
\affiliation{DTU Electro, Department of Electrical and Photonics Engineering, Technical University of Denmark, Ørsteds Plads 343, DK-2800 Kongens Lyngby, Denmark}

\author{Meng Xiong}
\affiliation{DTU Electro, Department of Electrical and Photonics Engineering, Technical University of Denmark, Ørsteds Plads 343, DK-2800 Kongens Lyngby, Denmark}
\affiliation{NanoPhoton - Center for Nanophotonics, Technical University of Denmark, Ørsteds Plads 345A, DK-2800 Kongens Lyngby, Denmark}

\author{Battulga Munkhbat}
\affiliation{DTU Electro, Department of Electrical and Photonics Engineering, Technical University of Denmark, Ørsteds Plads 343, DK-2800 Kongens Lyngby, Denmark}

\author{Elizaveta Semenova}
\affiliation{DTU Electro, Department of Electrical and Photonics Engineering, Technical University of Denmark, Ørsteds Plads 343, DK-2800 Kongens Lyngby, Denmark}
\affiliation{NanoPhoton - Center for Nanophotonics, Technical University of Denmark, Ørsteds Plads 345A, DK-2800 Kongens Lyngby, Denmark}

\keywords{quantum dots, site-controlled growth, fine-structure splitting, quantum communication, nanopyramids}

\begin{abstract}
Deterministic quantum light sources emitting at telecom wavelengths with vanishing fine-structure splitting (FSS) are essential components for scalable quantum communication.
While self-assembled Stranski-Krastanov (SK) quantum dots (QDs) are high-quality emitters, their random positioning and shape-induced anisotropy typically limit their use in entangled-photon applications.
In this work, we demonstrate site-controlled SK growth where InAs/InP QDs nucleate at the symmetric apexes of truncated InP nanopyramids.
Confining adatom diffusion to a small, symmetric nucleation area suppresses anisotropic growth, promoting the nucleation of highly symmetric QDs with FSS reduced to values below our statistically validated resolution limit of $9.2~\upmu$eV.
At the same time, lithographically defined nucleation sites enable deterministic control of the QD position, overcoming the limitations of conventional SK growth.
The high structural quality of single symmetric QDs is evidenced by the single-photon character of the emission ($g^{(2)}(0)=0.07^{+0.27}_{-0.07}$) spanning the S, C, and L telecom bands, with no evidence of lithography-induced defects affecting emission dynamics.
These results demonstrate that tailoring QD symmetry through nanopyramid growth engineering provides a route toward site-controlled emitters suitable for entangled photon generation and integrated quantum photonics devices.

\end{abstract}

\title{\papertitle{}}

\maketitle

\section*{Introduction} 
\vspace*{-3pt}

Quantum information technologies rely on the development of deterministic sources of non-classical light, particularly single photons and polarization-entangled photon pairs, for applications such as quantum communication, distributed quantum computing, and quantum sensing \cite{heindel_quantum_2023,sigov_quantum_2022,vajner_quantum_2022, tsukanov_exciton_2025}. Semiconductor quantum dots (QDs) are among the most promising solid-state platforms for this purpose \cite{chanana_ultra-low_2022,norman_perspective_2018, kim_hybrid_2017}, as their discrete energy levels enable the generation of single photons on demand with high brightness and stability \cite{valeri_generation_2024,holewa_solid-state_2025,zeuner_-demand_2021,wyborski_high-purity_nodate,ge_polarized_2024}. In particular, QDs emitting in the telecom wavelength range are highly attractive because existing optical fibers exhibit minimal transmission loss in this spectral region,  which is essential for scalable quantum communication networks. \cite{anderson_quantum_2020,vajner_quantum_2022,yu_telecom-band_2023, uppu_quantum-dot-based_2021}. Telecom-wavelength QDs can be achieved either via metamorphic growth on GaAs with strain relaxation through plastic deformation \cite{Semenova_2008_MetamorphicQD}, or through material systems that avoid strain-induced plastic relaxation and are therefore largely free from dislocation-related defects \cite{holewa_solid-state_2025}, such as InAs/InP \cite{Berdnikov2024NearCriticalSK}.
Furthermore, future quantum network architectures rely on the distribution of quantum correlations between distant nodes, making polarization-entangled photon sources a key enabling resource for quantum information processing. However, the generation of polarization-entangled photon pairs from QDs requires a vanishing exciton fine-structure splitting (FSS) \cite{plumhof_experimental_2012,lettner_strain-controlled_2021}. In conventional self-assembled Stranski–Krastanov (SK) QDs, anisotropic adatom diffusion and strain relaxation during QD nucleation lead to elongated dot shapes, resulting in large FSS values that significantly degrade entanglement fidelity \cite{seguin_size-dependent_2005,zeuner_-demand_2021,goldmann_excitonic_2013}. On InP platform, typical FSS values exceed $\SI{100}{\micro\electronvolt}$; exceptionally low FSS has been reported in specific cases through careful growth optimization \cite{kors_telecom_2018}. As an alternative to SK growth, droplet epitaxy (DE), which is not strain-driven, can produce more symmetric quantum dots, resulting in reduced FSS values on the order of $\SI{30}{\micro\electronvolt}$ \cite{holewa_droplet_2022, skiba-szymanska_universal_2017}. Despite these improvements, achieving vanishing FSS typically requires additional post-growth steps, such as applying external strain to fine-tune the QD symmetry and reach near-zero FSS. However, the tuning range is limited, and therefore an already low initial FSS (typically below $\SI{6}{\micro\electronvolt}$) is required to bring the FSS within the tunable range needed for high-fidelity polarization entanglement.\cite{lettner_strain-controlled_2021}. Consequently, overcoming this intrinsic asymmetry while maintaining high optical quality remains a major challenge for scalable QDs as telecom-band quantum light sources \cite{li_scalable_2023}.

Furthermore, precise knowledge of the QD location is essential for benefiting from light-matter interaction by integrating emitters with photonic structures, such as nanocavities, to enhance photon extraction efficiency and exploit the Purcell effect \cite{martin_purcell-enhanced_2025,kaupp_purcell-enhanced_2023}.
Several strategies have been developed to address this challenge \cite{mano_latticemismatched_nodate,deutsch_telecom_2023,chen_situ_2025,lettner_strain-controlled_2021, mccabe_techniques_2020}.
The typical approaches rely on fabricating photonic structures around randomly positioned QDs: either by statistical alignment, or by locating and pre-selecting individual QDs optically~\cite{sapienza_nanoscale_2015,dousse_controlled_2008, thon_strong_2009}.
While the latter enables deterministic nanodevice fabrication, relying on optical signals imposes fundamental resolution limits, and scalability remains a limiting factor for the development of advanced quantum networks \cite{holewa_high-throughput_2024, madigawa_assessing_2024}.

An alternative strategy is selective area epitaxy (SAE), in which lithographically patterned templates promote nucleation at specific sites by locally lowering surface energy \cite{baier_high_2004,felici_site-controlled_2009}.
Although this approach offers deterministic positioning, the high precision of the QD position can come at a cost of proximity to the fabrication-induced defects.
As a result, the optical quality of such QDs can be degraded by defects introduced during the template fabrication, as they act as non-radiative recombination centers that quench the emission \cite{clausen_determination_1989,fan_recent_2021, holewa_optical_2021}, or increase the spectral diffusion due to charge trapping. Addressing these limitations is crucial to realizing scalable, reliable quantum networks \cite{hou_review_2025}. Consequently, developing epitaxial growth techniques that control QD positioning and tailor morphology represents a key step toward achieving high-performance single-photon sources at telecom wavelengths \cite{mccabe_techniques_2020,zhao_site-controlled_2024,grose_development_2020}.

In this work, we demonstrate site-controlled InAs/InP QDs with low FSS nucleated at the apex of selectively grown InP nanopyramids, fabricated using e-beam lithography-defined SAE. The truncated apex of InP nanopyramids serves as well-defined, deterministic nucleation sites for the formation of InAs QDs\cite{poole_selective_2010, wong_controlled_2007,jung_selective_2022,zhou_submicron_2007}. This geometry spatially separates the quantum dot from etched interfaces by several hundred nanometers, mitigating surface-induced non-radiative recombination and preserving the optical quality of the emitter. Thus, the site-controlled InAs/InP QDs exhibit single-photon characteristics, with $g^{(2)}(0)=0.07^{+0.27}_{-0.07}$, with no evidence of lithography-induced degradation of the optical quality. 
The symmetric (001)-oriented apex, enclosed by other differently oriented crystallographic facets, restricts adatom diffusion to a small, well-defined nucleation site, suppressing anisotropic growth mechanisms that lead to QD elongation on planar surfaces and favoring the formation of QDs with improved in-plane symmetry and consequently reduced FSS below $\SI{9.2}{\micro\electronvolt}$. The strain pattern further locks QD nucleation at the apex center, minimizing positional misalignment with respect to the e-beam-defined site. 

The results presented demonstrate that nanopyramid-mediated growth offers a viable route toward site-controlled QDs with optical properties required for entangled-photon generation in scalable quantum photonic devices. 

\section*{Results and discussion}
 
\subsection{Engineering nucleation sites}    

\begin{figure*}
    \centering    
    \includegraphics[width=0.9\linewidth]{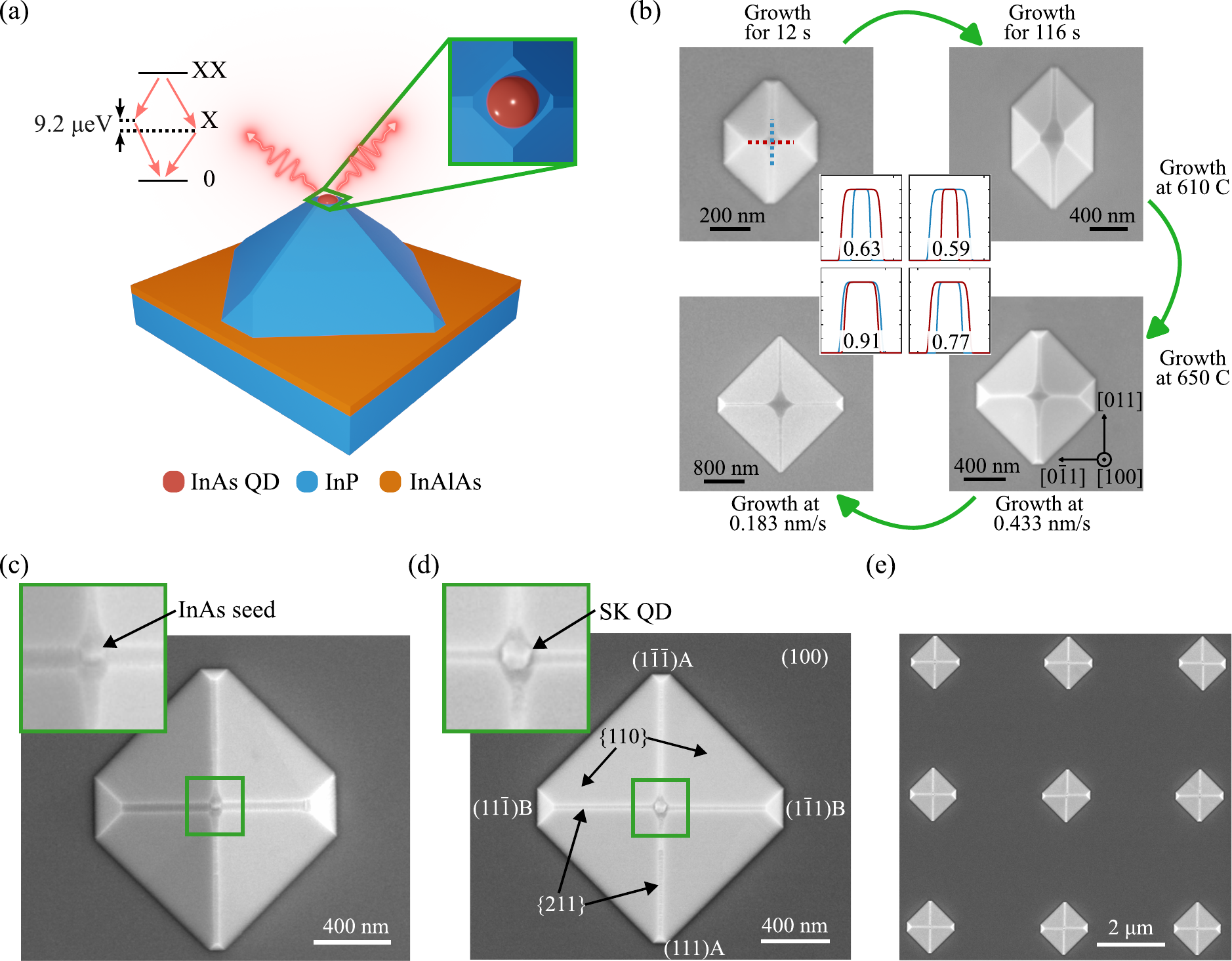}
    \caption{Epitaxial growth of QDs on top of nanopyramids.
    \figlab{a} Schematic of InAs QD as a single photon source on top of a truncated nanopyramid.
    \figlab{b} Symmetry optimization of truncated top, $\mathrm{FoM} = 1 - |L_x - L_y|/|L_x + L_y|$.Upper left: Pyramid grown for $\SI{12}{\second}$ at $\SI{610}{\degreeCelsius}$, truncated top is elongated in $[0 \bar{1}1]$. Upper right: Pyramid grown for $\SI{116}{\second}$ at $\SI{610}{\degreeCelsius}$, growth rate is suppressed for \{111\}B planes, hence truncated top is elongated in $[011]$. Bottom right: Temperature increase from $\SI{610}{\degreeCelsius}$ to $\SI{650}{\degreeCelsius}$, symmetry is improved, but elongation remains in $[0\bar{1}1]$. Bottom left: Reducing the growth rate from $\SI{0.433}{\nano\meter\per\second}$ to $\SI{0.183}{\nano\meter\per\second}$ produced a highly symmetrical pyramid top.
    \figlab{c} InP nanopyramid after As/P exchange to create a seed for QD growth -- it is optically inactive.
    \figlab{d} InP nanopyramid after As/P exchange and QDs growth. The zoomed-in image shows high symmetry of the QD.
    \figlab{e} Array of uniform, symmetric pyramids with SK QDs grown on top.}
    \label{fig:Summary}
\end{figure*}

Selective area growth occurs in the openings of a lithographically patterned hard mask, allowing growth within the openings while suppressing it elsewhere \cite{wang_shape_2019}.
Instead of a conventionally employed dielectric hard mask, such as silicon nitride (SiN) or silicon dioxide (SiO$_2$) \cite{dalacu_directed_2010,poole_selective_2010,wang_distribution_2009,yamakawa_cathodoluminescence_2002}, an epitaxially grown InAlAs layer serves as the mask material, which, upon exposure to air, forms a native oxide that is subsequently removed prior to regrowth, however, residual oxide species may remain, and the exposure process may additionally modify the surface energy, enabling a hard mask behavior while remaining compatible with subsequent regrowth processes. Openings are then patterned into the InAlAs layer by e-beam lithography, defining the nucleation sites for selective InP growth.
Additionally, using an epitaxial InAlAs layer with a matching thermal expansion coefficient eliminates issues with mask adhesion and mechanical stability at elevated temperatures, simplifying device integration and reducing fabrication complexity.
A schematic of the uncapped QD atop the pyramid is shown in \subfigref{Summary}{a}, with the pyramid positioned over the opening in InAlAs.

The pitch of the pyramids targets the favorable low QD density, indispensable for quantum photonic devices \cite{shang_ultra-low_2024, holewa_high-throughput_2024}.
Accordingly, the pitch was set as $\SI{4}{\micro\meter}$ and $\SI{5}{\micro\meter}$, corresponding to densities of $6.26\times10^6$ and $\SI[per-mode=power]{4e6}{\per\centi\meter\squared}$, respectively. While dielectric masks such as SiN or SiO$_2$ readily suppress parasitic growth between nucleation sites, it is challenging for InAlAs, as it does not completely inhibit residual material growth between predefined sites when the adatom diffusion length is shorter than half the pitch distance. 
Conversely, if the diffusion length is too long, pyramids may fail to nucleate in some openings.
To identify optimal growth conditions, the effects of temperature, growth rate, and V/III ratio were systematically studied, considering that the diffusion length increases with both temperature and V/III ratio \cite{hoglund_optimising_2006}.
These parameters must be finely tuned not only to achieve site-controlled growth but also to ensure uniformity across the nanopyramid array.

\subsubsection{Epitaxial growth of InP nanopyramids }

Selective growth of InP following the zinc-blende crystal structure on the patterned substrate leads to the formation of three-dimensional pyramids, as illustrated in \subfigref{Summary}{a}. During the nucleation stage, the effective growth rate on the (100) crystallographic plane dominates. As growth proceeds, the growth rates on the  \{110\} side facets and the (100) top facet become more balanced, causing the lateral dimensions of the truncated top facet to progressively decrease as the pyramid develops. Tailoring the growth parameters enables control of growth rates along different crystallographic directions, resulting in pyramidal structures with desired shapes. This facet-engineering approach enables precise modification of the top facet, which plays a key role in defining the in-plane dimensions and symmetry of the embedded QDs. To quantify apex symmetry, we define a figure of merit $\mathrm{FoM} = 1 - |L_x - L_y|/|L_x + L_y|$, where $L_x$ and $L_y$ are the lateral dimensions of the truncated top facet, with $\mathrm{FoM} = 1$ corresponding to a perfectly square apex.

As shown in \subfigref{Summary}{b} (top-left), the non-optimized pyramid exhibits elongation along the $[0\bar{1}1]$ direction.
However, when the InP growth time is increased from $\SI{12}{\second}$ to $\SI{116}{\second}$ at a growth temperature of $\SI{610}{\degreeCelsius}$, the elongation is observed along in the $[011]$ direction (\subfigref{Summary}{b}, top-right) .
This indicates that growth is suppressed by neighboring pyramids in the \{111\}B planes, while growth remains enhanced along the \{111\}A and \{110\} planes.
Increasing the temperature to $\SI{650}{\degreeCelsius}$ increases the diffusion length (\subfigref{Summary}{b}, bottom-right), which in turn reduces the growth rate along \{111\}A planes, while \{111\}B remains diffusion-limited by adjacent structures.
These B-type facets are group V-terminated, while A-type facets are group III-terminated.
As a result, the pyramid top becomes more symmetric, with the V/III ratio held constant at $143$.
Additional improvement was achieved by reducing the indium (In) precursor supply, thereby increasing the V/III ratio and lowering the overall growth rate, as shown in \subfigref{Summary}{b} (bottom-left). Additionally, the growth time was increased to $\SI{437}{\second}$ to produce pyramids of sufficient size to better assess the apex symmetry. This change reduces the kinetic preference for faster-growing facets, giving adatoms more time to diffuse and distribute more uniformly, which allows the system to approach a more thermodynamically balanced growth across different crystallographic directions.

Optimized growth conditions provide a baseline understanding of how pyramid geometry responds to different growth parameters. For QD nucleation, the growth rate was further reduced to $\SI{0.036}{\nano\meter\per\second}$ and the growth time shortened to $\SI{27.3}{\second}$, producing pyramids with an apex size matched to the expected QD size. These adjustments preserve the crystallographic balance achieved during optimization by maintaining the relative diffusion lengths across the \{111\}A, \{111\}B, and \{110\} facets, ensuring that apex symmetry is not compromised during the transition to QD growth conditions.

\subsubsection{Site-controlled QD nucleation} 

InAs preferentially nucleates at sites of lowest surface energy \cite{wong_controlled_2007}, typically near the pyramid base or on the InAlAs surface, rather than at the pyramid apex, even when the (100) top facet is exposed (see Supplementary Note 1). To promote QD apex-selective nucleation, a localized seed is formed prior to InAs growth via arsenic/phosphorus (As/P) exchange in an As-rich environment. During this growth interruption, In adatoms from the sidewall facets diffuse toward the pyramid apex, while the As–P exchange reconstructs the topmost monolayers \cite{Berdnikov2024NearCriticalSK}, forming a localized nucleus at the center of the top facet. This seed is optically inactive, as confirmed by cryogenic $\upmu$PL measurements, but serves as an effective seed for subsequent InAs QD growth.

The As/P exchange seed is highly sensitive to both As flux and temperature. Increasing either parameter enhances the efficiency of the exchange process and promotes the migration of atoms from high-index facets toward the pyramid apex. However, using an As flux that is too low results in nonuniform seed formation, while excessively high temperatures lead to unwanted residual growth on the \{211\} planes, see Supplementary Note 2. Notably, the QD growth rate must be significantly lower than that used for conventional Stranski-Krastanov (SK) QDs, due to the enhanced local growth rate arising from the limited number of available nucleation sites inherent to SAE. The final result, including the As/P exchange seed, the InAs QD growth, and the array of uniform, symmetric pyramids, is shown in \subfigsref{Summary}{c}{e}.
The zoomed-in SEM image in \subfigref{Summary}{d} shows a highly symmetric SK QD grown on the pyramid's apex.

\subsection{Low FSS in site-controlled QDs} 
  
To evaluate the symmetry of the QDs nucleated at the nanopyramid apex, we analyze their optical properties, focusing on their single-photon emission characteristics and exciton FSS. The dots were characterized by low temperature ($T = \SI{5}{\kelvin}$) micro-photoluminescence ($\upmu$PL), excitation power-dependent PL, time-resolved PL (TRPL), and polarization-dependent PL. The impact of the local QD growth environment was investigated by changing the pyramid height. QDs were grown on pyramids with heights of $\SI{515}{\nano\meter}$, $\SI{496}{\nano\meter}$ and $\SI{318}{\nano\meter}$. It is expected that a systematic reduction of the pyramid height can influence the FSS of the QD by enhancing its symmetry. Multiple QDs were measured for each pyramid height with the emission wavelength spanning from $\SI{1480}{\nano\meter}$ to $\SI{1580}{\nano\meter}$, covering the S, C, and L bands, which are highly relevant for long-haul optical communication systems or wavelength-division multiplexing (WDM) \cite{deng_challenges_2022, idris_wdmtdm_2016}.

\begin{figure}
    \centering
    \includegraphics[width=0.99\linewidth]{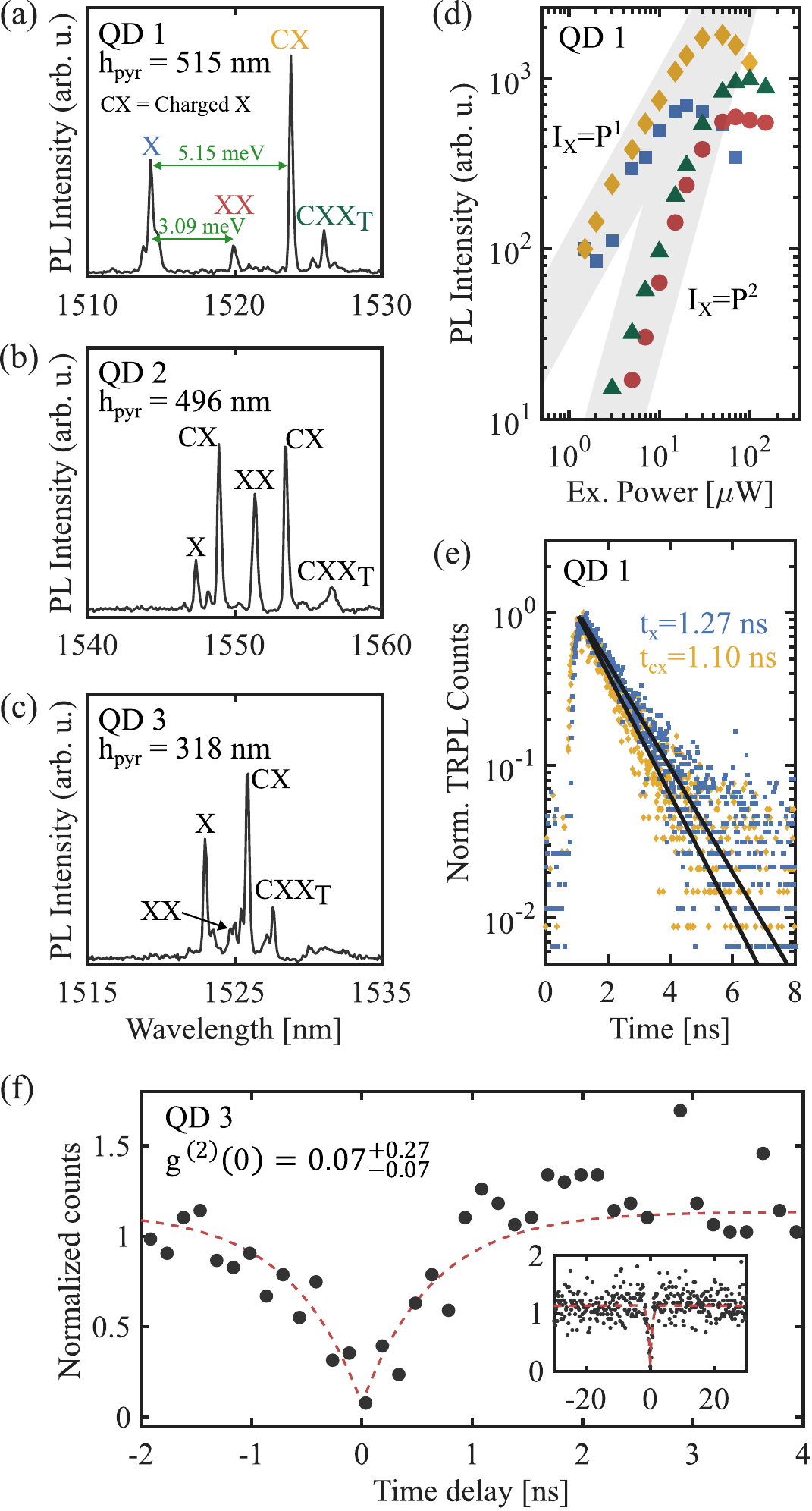}  
    \caption{Optical characterization of QDs on top of nanopyramids at $\SI{5}{\kelvin}$. 
    PL spectrum of \figlab{a} QD 1, \figlab{b} QD 2, and \figlab{c} QD 3 with assignment of identified excitonic complexes.
    $h_{\textrm{pyr}}$ corresponds to the pyramid height.
    \figlab{d}~$\upmu$PL intensity on excitation power with linear ($I_{xx}=P^1$) and quadratic ($I_{xx}=P^2$) dependency for QD 1. 
    \figlab{e} Time-resolved $\upmu$PL of exciton and charged exciton of QD 1.
    \figlab{f} Second-order autocorrelation function $g^{(2)}(\tau)$ of X from QD 3. 
    \label{fig:uPL}}
\end{figure}

In \subfigsref{uPL}{a}{c}, the $\upmu$PL spectra from QDs 1--3 are shown, which correspond to the pyramid height of $\SI{515}{\nano\meter}$, $\SI{496}{\nano\meter}$, and $\SI{318}{\nano\meter}$, respectively. The observed pattern of excitonic transitions is consistent with that of conventional SK InAs/InP QDs \cite{zielinski_excitonic_2020,holewa_bright_2022}: the neutral exciton (X) is found at the highest energy, followed by the biexciton (XX) and charged exciton (CX) at progressively lower energies. The biexciton and charged exciton binding energies in SK QDs are typically around $\SI{3}{\milli\electronvolt}$ and $\SI{5}{\milli\electronvolt}$~\cite{holewa_bright_2022,burakowski_heterogeneous_2024}, respectively, depending on the QD size and confinement potential. These values match those found in QD~1 (\subfigref{uPL}{a}), but QDs 2 and 3 exhibit lower binding energies while maintaining a similar emission wavelength. This behavior could be attributed to variations in the in-plane geometry of the QDs, while the out-of-plane geometry, which provides the dominant contribution to carrier confinement, remains largely unchanged.

\begin{figure*}
    \centering
    \includegraphics[width=0.8\linewidth]{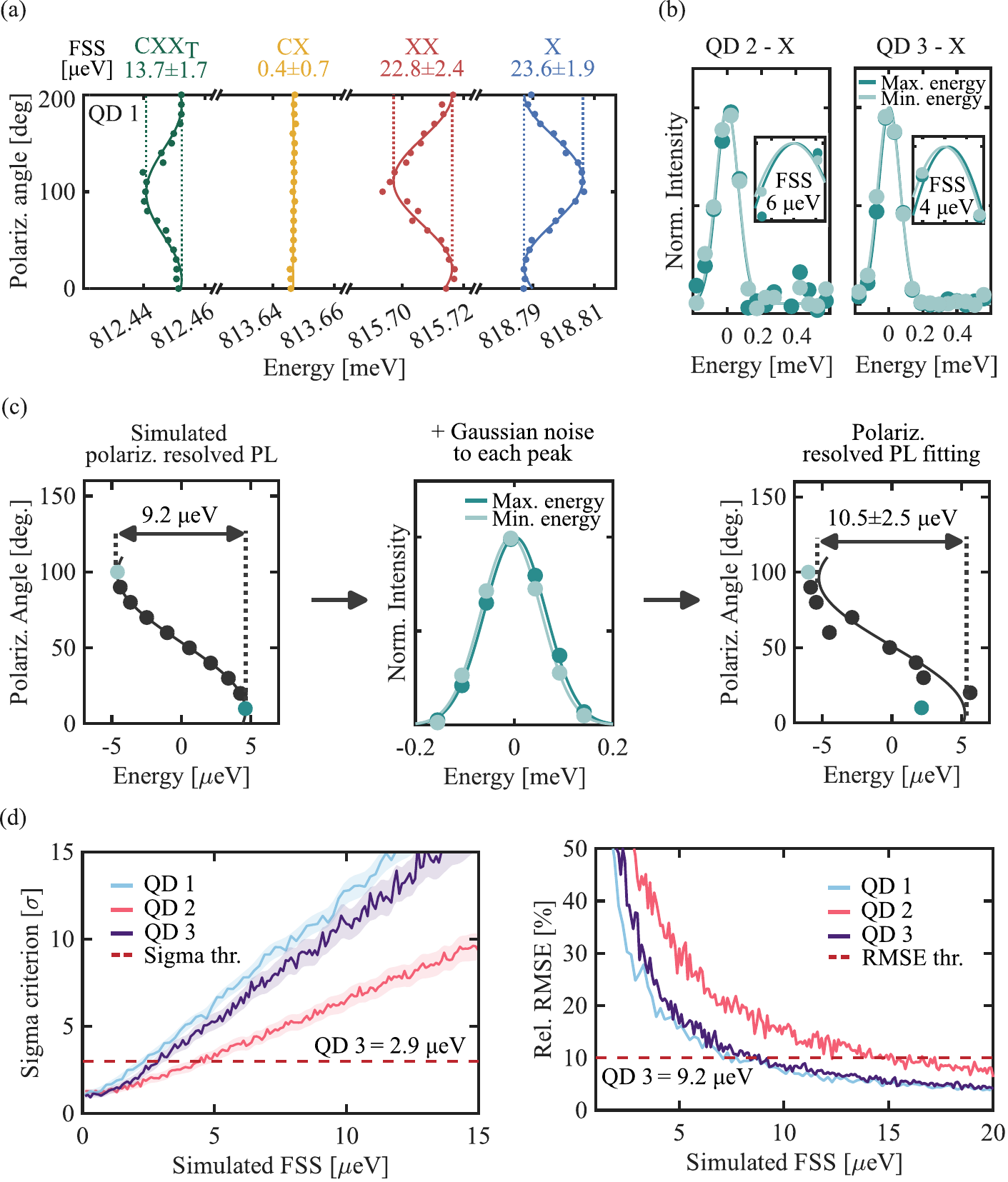}
    \caption{FSS extraction and numerical validation of experimental resolution limits and statistical accuracy.
    \figlab{a} Polarization dependence of emission energies for QD 1 with fitted FSS values for all excitonic complexes. 
    \figlab{b} Polarization-resolved $\upmu$PL spectra for X transition in QD 2 and QD 3 taken at maximum and minimum energy positions.   
    \figlab{c} Monte Carlo simulation workflow used to evaluate the FSS extraction limit. Simulated polarization-resolved spectra are generated using experimentally derived linewidths, intensities, and noise levels and analyzed using the same fitting procedure as the experimental data.
    \figlab{d} Determination of the minimum reliable FSS for QDs 1–3. The results are averaged over 100 Monte Carlo simulations for each simulated FSS value.
    Left: Statistical significance of the extracted FSS. A non-zero FSS is considered resolved when the fitted value exceeds three times its fitting uncertainty $\mu_{\mathrm{FSS}} > 3\delta_{\mathrm{FSS}}$ corresponding to a $3\sigma$ confidence level (dashed line). For QD 3, this criterion yields a minimum resolvable FSS of $\SI{2.9}{\micro\electronvolt}$.
    Right: Relative RMSE between the simulated and extracted FSS, quantifying the average percentage deviation of the FSS fit from the true value. The minimum accurately quantifiable FSS is defined as the value for which the RMSE falls below \SI{10}{\percent} (dashed line), corresponding to $\SI{9.2}{\micro\electronvolt}$ for QD 3.}
    
    \label{fig:Simulation}
\end{figure*}

Given the low QD densities of $\SI[per-mode=power]{6.26e6}{\per\centi\meter\squared}$ and $\SI[per-mode=power]{4e6}{\per\centi\meter\squared}$, together with a laser spot diameter of approximately $\SI{2}{\micro\meter}$, the emission spectra are expected to be from a single emitter at each measurement position. Consistent with this expectation, most spectra recorded show a single set of emission lines, confirming that the signal originates from a single QD at the pyramid apex.

The identification of the excitonic transitions was performed using excitation-power-dependent and TRPL measurements, with QD 1 data as a reference. As shown in \subfigref{uPL}{d}, two single-excitonic states are identified by their linear dependence on excitation power up to saturation. In addition, two transitions are observed, exhibiting an approximately quadratic power dependence characteristic of biexcitonic emission. TRPL measurements (\subfigref{uPL}{e}) show a decay time of $\SI{1.27}{\nano\second}$ for the neutral exciton, which is slightly shorter than typical values reported for random SK InAs/InP QDs \cite{holewa_bright_2022}. The charged exciton has a shorter lifetime of $\SI{1.10}{\nano\second}$, attributed to the presence of an additional carrier, which modifies the Coulomb interaction and increases the electron and hole wavefunctions overlap \cite{anderson_coherence_2021,katsumi_cmos-compatible_2022,ge_polarized_2024}.

\begin{table*}[!ht]
\caption{Summary of FSS for InP and GaAs-based QDs.}
\centering
\begin{tabular}{lcccc}
\toprule 
\textbf{Method} & \textbf{Platform} & \textbf{FSS [$\mu$eV]} & \textbf{Wavelength [$\mu$m]}
& \textbf{Reference} \\
\midrule 
InAsP nanowire QDs     & InP    & $4.6$   & 1.26   & \onlinecite{alqedra_entangled_2025} \\
InAs droplet epitaxy     & InP(001)    & $29\pm1$    & 1.55   & \onlinecite{skiba-szymanska_universal_2017} \\
InAs QDs     & InP(001)    & $25\pm4$    & 1.54-1.65   & \onlinecite{wakileh_approaching_2025} \\
InAs QDs     & InP(100)    & $2$    & 1.54-1.56   & \onlinecite{kors_telecom_2018} \\
InAs droplet epitaxy     & GaAs(111)A    & $16\pm6$    & 1.3   & \onlinecite{tuktamyshev_telecom-wavelength_2021} \\
InGaAs QDs on pillars     & GaAs(111)B   & $1.6\pm0.8$    & 0.86   & \onlinecite{juska_self-aligned_nodate} \\
InAs droplet epitaxy     & GaAs    & $31\pm1$    & 0.9   & \onlinecite{skiba-szymanska_universal_2017} \\
InGaAs QDs     & GaAs    & $56\pm5$    & 1.2   & \onlinecite{kettler_neutral_2016} \\
InAs QDs on nanopyramids     & InP    & $<9.2$    & 1.514   & This approach \\
\bottomrule 
\end{tabular}
\label{tab:SummaryFSS}
\end{table*}

The single-photon emission nature of the QD was verified through the second-order correlation function $g^{(2)}(0)$ measured on the exciton line using a Hanbury–Brown and Twiss setup for QD 3. The coincidence histogram shows almost no events at zero time delay, yielding $g^{(2)}(0)=0.07^{+0.27}_{-0.07}$ after normalization. Due to the limited out-of-plane emission and the absence of a distributed Bragg reflector (DBR) or metallic mirror at the bottom of the stack to enhance photon extraction, the collection efficiency is low. Nevertheless, the clear anti-bunching behavior observed in \subfigref{uPL}{f} confirms that the emitter operates as a single-photon source.

\subsubsection{Polarization-dependent measurement} 
Polarization-dependent PL measurements were performed to confirm the charge state of the observed transitions; the results are summarized in \subfigsref{Simulation}{a}{b}. Due to the resident intrinsic anisotropy of the QD, the exciton FSS is observed in $\upmu$PL as splitting of X and XX lines into a doublet with orthogonal linear polarizations (H and V).
However, due to the monochromator's limited spectral resolution at telecom wavelengths and the observed emission linewidths, the FSS doublet cannot be spectrally resolved.
Instead, the X and XX transitions appear as single broadened peaks.
As shown in \subfigref{Simulation}{a}, two transitions (marked with blue and red) exhibit anti-phase polarization-dependent oscillations with amplitudes of $\SI{23.6(1.9)}{\micro\electronvolt}$ and $\SI{22.8(2.4)}{\micro\electronvolt}$, respectively, which is characteristic of a biexciton–exciton cascade.

The extracted FSS for QD 1 is close to the typical values reported for optimized InAs QDs grown by standard random methods such as SK or droplet epitaxy on InP, as summarized in \tabref{SummaryFSS}, which also includes additional references for the GaAs platform. 
Although this value lies below the monochromator's nominal resolution, it can still be resolved when the emission intensity is sufficiently high and the spectral linewidth is narrow enough, as shown in the next section.
The positive biexciton binding energy shifts the XX transition line to lower energy than the X line.
The charged exciton (CX$^{+/-}$, \subfigref{Simulation}{a}, yellow) shows no detectable splitting with an extracted FSS close to 0, as expected.
In contrast, the last resolved peak (marked in green) exhibits polarization oscillations that are in phase with the biexciton but with a smaller FSS.
Given its quadratic power dependence (\subfigref{uPL}{d}), this peak can be attributed to a charged biexciton (CXX$_{\textrm{T}}$), as reported in previous analysis \cite{kettler_neutral_2016}.

QD 2 and 3 show similar spectral features as QD 1, with the addition of a charged exciton transition (CX$^{+/-}$) located between the neutral exciton and biexciton. The second CX$^{+/-}$ line likely arises from variations in the local charge environment.
Furthermore, as shown in \subfigref{Simulation}{b}, the FSS decreases systematically with decreasing nanopyramid height, reaching values as low as $\SI{4}{\micro\electronvolt}$. This trend is consistent with the improved symmetry of the truncated pyramid top resulting from enhanced control over the local growth environment, which suppresses the anisotropic elongation of the QDs.
The peaks at maximum and minimum of the polarization-dependence measurement highlight the challenge of resolving such small FSS values, as the splitting is significantly lower than the emission linewidth.
In addition, the limited spectral resolution of the optical setup (approximately $\SI{70}{\micro\electronvolt}$) and background noise further complicate the extraction of the FSS.

\begin{figure*}
    \centering
    \includegraphics[width=0.9\linewidth]{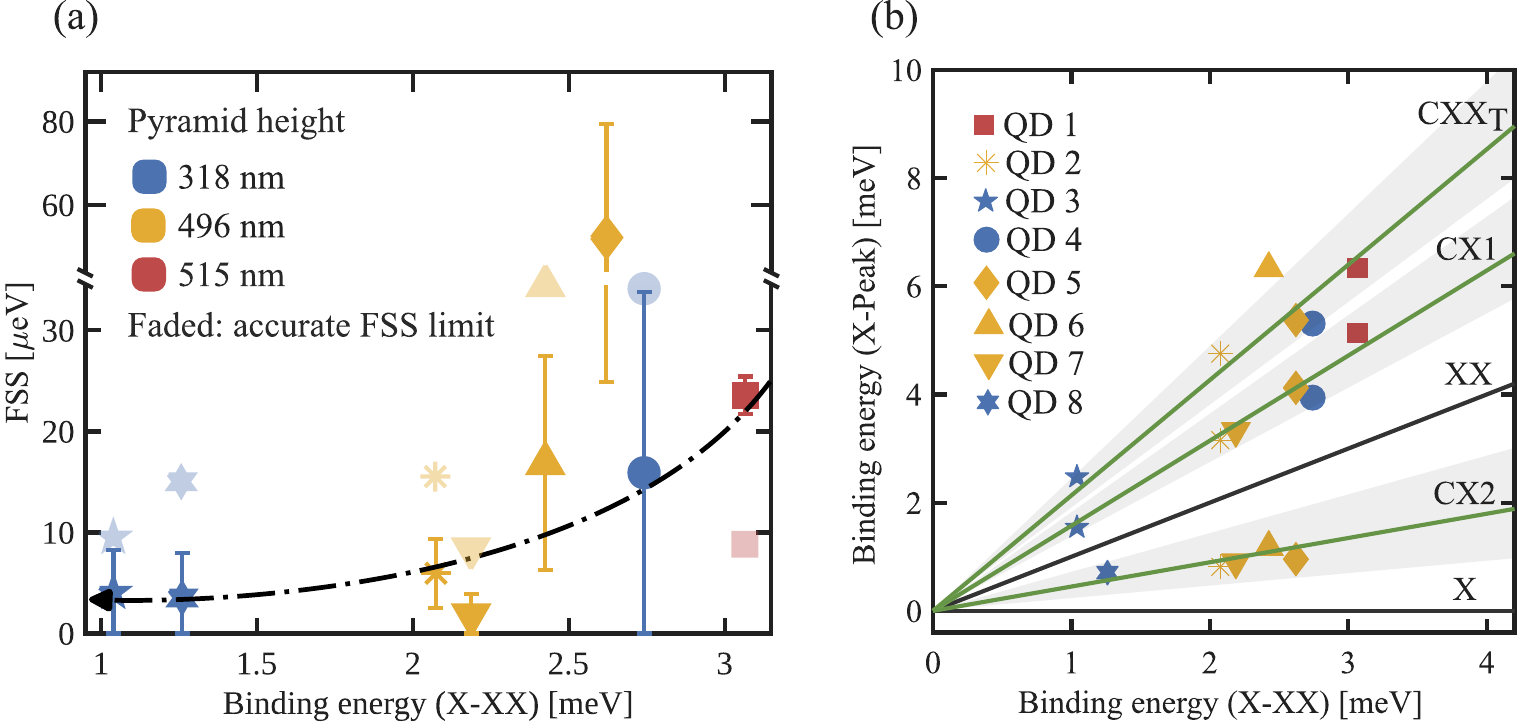}
    \caption{Dependence of excitonic complex properties on pyramid height.
    \figlab{a} FSS reduction induced by decreasing pyramid height and the corresponding X–XX binding energy.
    Full color -- experimentally determined FSS values.
    Faded symbols -- simulation-based, minimal accurately quantifiable splitting, defined by \eqnref{RMSE}.
    \figlab{b} Binding energy reduction of all excitonic complexes with decreasing pyramid height.}
    \label{fig:Binding_energies}
\end{figure*}

\subsubsection{Modeling of minimum resolvable FSS} 

Because the observed fine-structure splitting is an order of magnitude smaller than the emission linewidth, traditional fitting may be susceptible to experimental noise. To establish a rigorous lower bound for what our setup can reliably distinguish from zero, we employed Monte Carlo simulations to quantify our statistical detection limit.

To determine the minimum FSS that can be reliably measured under the present experimental conditions, Monte Carlo simulations were performed to reproduce the polarization-resolved measurements of QDs 1–3. The simulations account for the finite spectral resolution, emission linewidth, detector sampling, and noise, enabling a quantitative assessment of the accuracy and precision of the extracted FSS values.

For each QD, the exciton emission was modeled as a single Gaussian peak with the experimentally extracted linewidth, intensity, and spectrum noise. The effect of FSS was included by introducing a sinusoidal shift of the peak center as a function of the polarization angle. Each simulated spectrum was discretized according to the detector's pixel size. The polarization sampling matched the experimental angles measured with angles ranging from \ang{0} to \ang{200} in \ang{10} steps for QD 1 and 2, and from \ang{0} to \ang{100} in \ang{10} steps for QD 3, as shown in \subfigref{Simulation}{c}, left. For each polarization angle, detector noise was added to the simulated peak based on the noise level extracted from the corresponding experimental $\upmu$PL spectra (\subfigref{Simulation}{c}, middle). The resulting datasets were analyzed using the same peak-fitting routine applied for the experimental measurements to extract the FSS, yielding an extracted FSS value for comparison with the known input FSS (\subfigref{Simulation}{c}, right). The simulated FSS was swept from $\SI{20}{\micro\electronvolt}$ down to $\SI{0}{\micro\electronvolt}$, and for each value, 100 independent simulations were performed to statistically evaluate the accuracy and robustness of the extraction procedure.

Two complementary metrics were used. The first evaluates whether a non-zero FSS can be statistically distinguished from zero, while the second evaluates whether the magnitude of the FSS can be accurately quantified.
To establish the detection limit, a fitted FSS was considered resolved when its value exceeded three times the associated fitting uncertainty, $\mu_{\mathrm{FSS}} > 3\delta_{\mathrm{FSS}},$ where $\mu_\text{FSS}$ is the fitted FSS value and $\delta_\text{FSS}$ its fit uncertainty. This corresponds to a $3\sigma$ confidence level (\SI{99.7}{\percent}). Applying this criterion to the Monte Carlo simulations gives a minimum resolvable FSS of \SI{2.9}{\micro\electronvolt} for QD~3 (\subfigref{Simulation}{d}, left).

The accuracy of the extracted FSS was evaluated using the relative root mean square error (RMSE), defined as
\begin{equation}\label{eq:RMSE}
\mathrm{RMSE} = 
\sqrt{\frac{1}{N} \sum_{i=1}^{N} 
\left( \frac{\hat{\Delta}_i - \Delta_i}{\Delta_i} \right)^2} \times 100,
\end{equation}
where $\hat{\Delta}_i$, $\Delta_i$ and $N$ denote the extracted FSS value from the fit, the simulated true FSS and the number of trials, respectively. The RMSE quantifies the average relative deviation of the extracted FSS from the true input value. In this work, an RMSE below \SI{10}{\percent} was adopted as the criterion for accurate quantification of the FSS, serving as a practical threshold that ensures the uncertainty remains substantially smaller than the observed differences between samples.
Under current experimental conditions, this criterion establishes a threshold of \SI{9.2}{\micro\electronvolt} for QD 3 (\subfigref{Simulation}{d}, right).

The Monte Carlo analysis gives two complementary thresholds for QD 3. The detection limit ($\SI{2.9}{\micro\electronvolt}$) defines the minimum FSS distinguishable from zero at $3\sigma$ confidence. The measured FSS of $\SI{4}{\micro\electronvolt}$ satisfies this criterion and is therefore statistically nonzero. The quantification limit ($\SI{9.2}{\micro\electronvolt}$) defines the minimum FSS whose magnitude can be extracted with less than \SI{10}{\percent} relative error under the present experimental conditions. Since the measured FSS of QD 3 falls between these limits, we can conclude that the FSS is nonzero and below $\SI{9.2}{\micro\electronvolt}$ and this is the value adopted throughout this work.

\subsubsection{Correlation of FSS with binding energies of excitonic complexes}

Excitonic binding energies are fundamentally determined by the direct Coulomb integrals between carrier wavefunctions, modified significantly by many-body correlation effects, whereas the fine-structure splitting (FSS) of the neutral exciton is induced by the anisotropic part of the electron-hole exchange interaction.
While the excitonic binding energies and FSS originate from different physical interactions, both quantities depend on the spatial distribution and overlap of the electron and hole wavefunctions and may therefore evolve concurrently as the confinement potential becomes more symmetric.
Therefore, a correlated evolution of binding energies and FSS is expected as the nanostructure's in-plane symmetry is modified~\cite{Mrowinski2016,Sarkar2006,Seguin2007}.

Additional QDs were measured for the pyramids with a height of $\SI{496}{\nano\meter}$ and $\SI{318}{\nano\meter}$ to study the evolution of the FSS and excitonic binding energies.
The growth conditions and optical characterization are summarized in \tabref{Summary-pyramids-height}.
As a first observation, no clear correlation is found between emission wavelength and nanopyramid height, indicating that the dominant confinement energy of the QDs is not significantly affected by this parameter.
In contrast, the FSS decreases with decreasing pyramid height, reaching lower values for pyramids with a height of $\SI{318}{\nano\meter}$ (\subfigref{Binding_energies}{a}). This trend is attributed to the more symmetric apex of smaller pyramids, which reduces the in-plane asymmetry of the QD and consequently lowers the FSS.
A systematic evolution of the excitonic binding energies is observed, with all excitonic transitions moving closer together as the pyramid height is reduced (\subfigref{Binding_energies}{b}).

This trend reflects a proportional reduction of the binding energies of the excitonic complexes, suggesting a modification of the direct Coulomb integrals between confined carriers \cite{kadantsev_theory_2010,mekni_fine_2021, van_venrooij_fine_2024}. 
Notably, while the binding energies of the observed complexes scale linearly with the X-XX binding energy, the FSS exhibits a more pronounced, non-linear sensitivity to these structural changes. This is consistent with its high sensitivity to both the electron–hole exchange interaction and the restoration of in-plane symmetry.
The observed correlation between the reduction in FSS and the shift in excitonic binding energies suggests a common physical origin: the transition toward a more isotropic confinement potential, which simultaneously minimizes the anisotropic exchange interaction and reconfigures the mutual position and spatial overlap of the carrier wavefunctions~\cite{Mrowinski2016,Sarkar2006,Seguin2007}.

While the buried QD morphology cannot be directly resolved in the present structures, the systematic reduction of FSS together with the evolution of excitonic binding energies strongly suggests that the enhanced symmetry of the nanopyramid apex is transferred to the confinement potential of the QDs.
Importantly, the observed reduction in the FSS is achieved without compromising the emission wavelength, highlighting the pyramid height as an effective tuning parameter for QD symmetry and, ultimately, for controlling the FSS.

\begin{table*}[!ht]
\caption{Summary of growth conditions for InP nanopyramids, QD $\upmu$PL emission wavelengths, and minimum resolvable FSS from simulations validated against experimental measurements.}
\centering
\setlength{\tabcolsep}{10pt} 
\begin{tabular}{l S[table-format=1.3] S[table-format=2.1] S[table-format=3.0] S[table-format=1.0] S[table-format=4.0] S[table-format=2.1] S[table-format=2.1]}
\toprule 
\multicolumn{5}{c}{\textbf{Pyramid growth conditions}} & \multicolumn{3}{c}{\textbf{QD}} \\ 
\cmidrule[1.2pt](lr){1-5} \cmidrule[1.2pt](lr){6-8} 
\addlinespace[0.5ex]
\textbf{ID} & {\textbf{Growth rate}} & {\textbf{Growth time}} & {\textbf{Height}} & {\textbf{Pitch}} & {\textbf{$\lambda$ - X}} & {\textbf{Measured FSS}} & {\textbf{Resolvable FSS}} \\
& {[nm/s]} & {[s]} & {[nm]} & {[$\si{\micro\meter}$]} & {[nm]} & {[$\si{\micro\electronvolt}$]} & {[$\si{\micro\electronvolt}$]}\\
\midrule 
QD 1    & 0.186    & 43.7 & 515 & 5 & 1514 & 23.6   & 8.9\\
QD 2    & 0.038    & 27.3 & 318 & 5 & 1549 & 6.0    & 15.3\\
QD 3    & 0.186    & 43.7 & 496 & 4 & 1523 & 4.0    & 9.2\\
QD 4    & 0.186    & 43.7 & 496 & 4 & 1568 & 15.9   & 44.4\\
QD 5    & 0.186    & 43.7 & 496 & 4 & 1543 & 52.2   & 170.1\\
QD 6    & 0.186    & 43.7 & 496 & 4 & 1480 & 16.8   & 46.0\\
QD 7$^{\ast}$    & 0.038    & 27.3 & 318 & 5 & 1511 & 1.8    & 8.2\\
QD 8    & 0.038    & 27.3 & 318 & 5 & 1520 & 3.5    & 15.0\\
\bottomrule 
\end{tabular}
\par\smallskip
{\raggedright\footnotesize $^{\ast}$ QD 7 is not considered the QD with the lowest FSS since the measured value is below the $3\sigma$ criterion.\par}
\label{tab:Summary-pyramids-height}
\end{table*}

\section*{Conclusions}
We have demonstrated the fabrication of site-controlled InAs/InP QDs grown by MOVPE at the apexes of lithographically defined truncated nanopyramids. By exploiting confined adatom diffusion at the symmetric pyramid apex and the As/P exchange-seeding process, site-controlled QD nucleation is achieved. Systematic optimization of nanopyramid growth parameters enables precise tuning of apex size and symmetry, directly reducing the FSS to below $\SI{9.2}{\micro\electronvolt}$, indicating a promising route toward deterministic entangled-photon sources.

The QDs exhibit telecom-band emission spanning the S-, C-, and L-bands, with single-photon purity confirmed by Hanbury–Brown and Twiss measurements yielding $g^{(2)}(0)=0.07^{+0.27}_{-0.07}$, with no evidence of lithography-induced degradation of optical quality. Polarization-dependent PL measurements reveal a consistent reduction in FSS as pyramid size decreases. This observation aligns with the spectral binding-energy trend, in which all transitions move closer together as the FSS decreases, as expected for an ideal system.

The demonstrated platform combines site-controlled QDs with telecom-band emission and low FSS in a straightforward, scalable approach, making it a promising candidate for integration with photonic cavities and quantum photonic circuits.
Although the measured FSS values remain above the threshold required for direct observation of high-fidelity polarization entanglement, the demonstrated growth strategy establishes a scalable route toward further symmetry engineering of telecom-wavelength QDs.
Future work will also focus on Purcell-enhanced emission through cavity coupling to improve photon extraction efficiency and on further optical characterization to assess photon indistinguishability and coherence for quantum network applications.

\section*{Methods}
\setcounter{subsection}{0}
\subsection{Fabrication}
The samples were grown by low-pressure MOVPE using a TurboDisc reactor\textsuperscript{\textregistered} on a (100)-oriented InP wafer using trimethylindium (TMIn), trimethylgallium (TMGa), phosphine (PH$_3$), tertiarybutylphosphine (TBP), and arsine (AsH$_3$) as precursor sources. H$_2$ was used as carrier gas. A $\SI{250}{\nano\meter}$ InP buffer layer and a 14~nm InAlAs layer were first grown to form a clean and uniform surface. Nanoholes were then defined in the InAlAs layer using JEOL JBX9500FSZ electron-beam writer at 100~kV, serving as nucleation sites for the growth of InP nanopyramids. The InAlAs layer was dry etched using Inductively Coupled Plasma (ICP). 

Prior to growth, the samples underwent a cleaning procedure comprising an oxygen-plasma ashing step, followed by immersion in a solution of ammonium hydroxide (NH$_4$OH) and deionized water for $\SI{1}{\minute}$ to remove impurities and the native InP oxide. The samples were then loaded into the MOVPE reactor and thermally deoxidized at $\SI{650}{\degreeCelsius}$ to ensure complete oxide removal. The InP pyramids were grown at the same temperature.

For QDs growth, the optimal growth conditions were found to be an InP pyramid height of $\SI{515}{\nano\meter}$, $\SI{496}{\nano\meter}$ and $\SI{318}{\nano\meter}$, with the As/P exchange performed at $\SI{470}{\degreeCelsius}$ and an AsH$_3$ flux of $\SI{440}{\micro\mole\per\minute}$. The InAs QDs were subsequently grown at a rate of $\approx 0.12~\mathrm{ML\,s^{-1}}$, a V/III ratio of 132, TMIn flux of $\SI{3.33}{\micro\mole\per\minute}$, and AsH$_3$ flux was kept at the same value as the As/P step. The samples for optical investigations were subsequently overgrown with an initial InP cap of $\SI{10}{\nano\meter}$ at $\SI{470}{\degreeCelsius}$ and later completed with $\SI{100}{\nano\meter}$ at $\SI{610}{\degreeCelsius}$.

\subsection{Optical measurements}
All optical measurements were performed using a low-temperature micro-photoluminescence (\textmu PL) setup based on a closed-cycle cryostat (attoDRY800xs, Attocube) operating at a base temperature of 4 K. The cryostat is equipped with piezoelectric nanopositioners and a microscope objective (NA = 0.8, LT-APO/Telecom Attocube objective). Excitation was performed with an above-band 650 nm semiconductor diode laser (LDH-D-C-650, PicoQuant) operating in CW and pulsed modes, with below 100 ps pulses at an $\SI{80}{\mega\hertz}$ repetition rate. PL characterization was based on a 0.328 m focal-length monochromator (Kymera 328i, gratings: 150, 600 lines/mm, Andor, Oxford Instruments) with (In, Ga)As a linear array detector (iDus, Andor, Oxford Instruments) with the maximum resolution of the setup up to $\sim$0.07 meV. The polarization-resolved PL characterizations were obtained using a pair of a half-wave plate and a linear polarizer. Time-correlated single-photon-counting mode was used to perform time-resolved PL characterization and single-photon statistics measurements. As a detection setup, we used a single-photon counting module (Time Tagger Ultra, Swabian Instruments) together with superconducting nanowire single-photon detectors (ID281 SNSPD, ID Quantique). 

\vspace*{2mm}
\noindent{\bf References}

\bibliography{references}

\vspace*{2mm}
\noindent{\bf \normalsize Acknowledgment}\\
 The authors are thankful to Shima Kadkhodazadeh and Mikhail Nestoklon for their valuable suggestions regarding QD characterization and data analysis.  \\

\noindent{\bf \normalsize Funding}\\
The authors acknowledge financial support from the Danish National Research Foundation through NanoPhoton -- Center for Nanophotonics, grant number DNRF147, and the QPIC1550 project, funded by the EU Horizon Europe program under Grant Agreement no. 101135785.
P.\,H. was supported by the Polish National Agency for Academic Exchange under the Polish Returns program.
P.\,W. and B.\,M. acknowledge support from the European Research Council (ERC-StG ``TuneTMD'', grant no. 101076437),  Villum Fonden (project no. VIL53033), and European Research Council (ERC-CoG ``Unity'', grant no. 865230).\\

\noindent{\bf \normalsize Author Contributions}\\
ES, CR, and PH conceived the idea and designed the experiment. CR fabricated the samples, including epitaxial growth and further nanofabrication. 
PW and BM performed the optical characterization of quantum dots.
CR and PH analyzed and interpreted all the optical data.
CR, under the lead of PH, developed the FSS simulation model.
ES, PH, and MX supervised and coordinated the project. All authors discussed the results.
CR, PH, and ES contributed to manuscript writing, with input from all co-authors. All authors approved the final version of the manuscript.\\

\noindent{\bf \normalsize Disclosures}\\
The authors declare no conflicts of interest.\\

\end{document}